\documentclass{article}
\usepackage[utf8]{inputenc}
\usepackage{graphicx}
\usepackage{cite}

\begin{document}
%

\title{Sentiment Aggregate Functions for Political Opinion Polling using Microblog Streams}

\author{Pedro Saleiro$^{1,2,3}$, Lu\'{i}s Gomes$^{1,2}$ and Carlos Soares$^{1,2, 4}$\\
DEI-FEUP$^{1}$, Labs Sapo UP $^{2}$, LIACC$^{3}$, INESC-TEC$^{4}$,\\University of Porto,\\Rua Dr. Roberto Frias, s/n, Porto, Portugal}

\maketitle
\begin{abstract}
The automatic content analysis of mass media in the social sciences has become necessary and possible with the raise of social media and computational power. One particularly promising avenue of research concerns the use of sentiment analysis in microblog streams. However, one of the main challenges consists in aggregating sentiment polarity in a timely fashion that can be fed to the prediction method. We investigated a large set of sentiment aggregate functions and performed a regression analysis using political opinion polls as gold standard. Our dataset contains nearly 233 000 tweets, classified according to their polarity (positive, negative or neutral), regarding the five main Portuguese political leaders during the Portuguese bailout (2011-2014). Results show that different sentiment aggregate functions exhibit different feature importance over time while the error keeps almost unchanged.

\end{abstract}


\section{Introduction}\label{sec:intro}
Surveys and polls using the telephone are widely used to provide information of what people think about parties or political personalities\cite{JSS}. Surveys randomly select the electorate sample, avoiding selection bias, and are designed to collect the perception of a population regarding some subject, such as in politics or marketing. However this method is expensive and time consuming \cite{Connor,JSS}. Furthermore, over the years it is becoming more difficult to contact people and persuade them to participate in these surveys \cite{Kohut}.

On the other hand, online publication of news articles is a standard behavior of news outlets and the raise of social media, namely Twitter and Facebook, has changed the way people interact with news \cite{Bermingham,Saleiro2016}. This way, people are able to react and comment any news in real time. One challenge that several research works have been trying to solve is to understand how opinions expressed on social media, and their sentiment, can be a leading indicator of public opinion. 
However, at the same time there might exist simultaneously positive, negative and neutral opinions regarding the same subject. Thus, we need to obtain a value that reflects the general image of each political target in social media, for a given time period. To that end, we use sentiment aggregate functions. In summary, a sentiment aggregate function calculates a global value based on the number of positive, negative, and neutral mentions of each political target, in a given period. We conducted an exhaustive study and collected and implemented several sentiment aggregate functions from the state of the art \cite{Bermingham, pred12, pred6, pred3, livne2011party, tumasjan2010predicting, Gayo-Avello2012, O'Connor2010, Chung2011}.

Thus, the main objective of our work is to study and define a methodology capable of successfully estimating the polls results, based on opinions expressed on social media, represented by sentiment aggregators. We applied this problem to the Portuguese bailout case study, using Tweets from a sample of the Portuguese Tweetosphere and Portuguese polls as gold standard. Given the monthly periodicity of polls, we needed to monthly aggregate data. This approach allows each aggregator value to represent the monthly sentiment for each political party. Due to the absence of a general sentiment aggregate function suitable for different case studies, we decided to include all aggregate functions as features of the regression model. Therefore the learning algorithm is able to adapt to the most informative aggregate functions through time.

In the next Section we review related work. In Section \ref{sec:methodology} we present the methodology we implemented. We describe data in Section \ref{sec:data} followed by the experimental setup in Section \ref{sec:experimental}. In Section \ref{sec:results} we present and discuss the results we obtained, while Section \ref{sec:conclusions} is reserved for some conclusions taken from our study, and for future work.

\section{Related Work}\label{sec:related}
Content analysis of mass media has an established tradition in the social sciences, particularly in the study of effects of media messages,
encompassing topics as diverse as those addressed in seminal studies of newspaper editorials \cite{lasswell1952}, media agenda-setting
\cite{mccombs1972}, or the uses of political rhetoric \cite{moen1990}, among many others. By 1997, Riffe and Freitag \cite{riffe1997}, 
reported an increase in the use of content analysis in communication research and suggested that digital text and computerized means for its
extraction and analysis would reinforce such trend. Their expectation has been fulfilled: the use of automated content analysis has by now surpassed the use of hand coding \cite{neuendorf2002}. The increase in the digital sources of text, on the one hand, and current advances in
computation power and design, on the other, are making this development both necessary and possible, while also raising awareness about the
inferential pitfalls involved \cite{hopkins2010, chall2}.

One avenue of research that has been explored in recent years concerns the use of social media to predict present and future political events, namely electoral results \cite{Bermingham, pred12, pred6, pred3, livne2011party, tumasjan2010predicting, Gayo-Avello2012, O'Connor2010, Chung2011}. Although there is no consensus about methods and their consistency \cite{Metaxas2011, Gayo-Avello2011}. Gayo-Avello \cite{pred_survey} summarizes the differences between studies conducted so far by stating that they vary about period and method of data collection, data cleansing and pre-processing techniques, prediction approach and performance evaluation. One particular challenge when using sentiment is how to aggregate opinions in a timely fashion that can be fed to the prediction method. Two main strategies have been used to predict elections: buzz, i.e., number of tweets mentioning a given candidate or party and the use of sentiment polarity. Different computational approaches have been explored to process sentiment in text, namely machine learning and linguistic based methods \cite{pang2008, Kouloumpis2011, nakov2013}. In practice, algorithms often combine both strategies. 

 Johnson et al. \cite{JSS} concluded that more than predicting elections, social media can be used to gauge sentiment about specific events, such as political news or speeches. Defending the same idea, Diakopoulos el al. \cite{Diakopoulos} studied the global sentiment variation based on Twitter messages of an Obama vs McCain political TV debate while it was still happening. Tumasjan et al. \cite{tumasjan2010predicting} used Twitter data to predict the 2009 Federal Election in Germany. They stated that ``the mere number of party mentions accurately reflects the election result''. Bermingham et al. \cite{Bermingham} correctly predicted the 2011 Irish General Elections also using Twitter data. Gayo-Avello et al. \cite{Gayo-Avello2011} also tested the share of volume as predictor in the 2010 US Senate special election in Massachusetts.
On the other hand, several other studies use sentiment as a polls result indicator. Connor et al. \cite{O'Connor2010} used a sentiment aggregate function to study the relationship between the sentiment extracted from Twitter messages and polls results. They defined the sentiment aggregate function as the ratio between the positive and negative messages referring an specific political target. 
They used the sentiment aggregate function as predictive feature in the regression model, achieving a correlation of 0.80 between the results and the poll results, capturing the important large-scale trends. Bermingham et al. \cite{Bermingham} also included in their regression model sentiment features. Bermingham et al. introduced two novel sentiment aggregate functions. For inter-party sentiment, they modified the share of volume function to represent the share of positive and negative volume. For intra-party sentiment , they used a log ratio between the number of positive and negative mentions of a given party. Moreover, they concluded that the inclusion of sentiment features augmented the effectiveness of their model. 

Gayo-Avello et al. \cite{Gayo-Avello2011} introduced a different aggregate function. In a two-party race, all negative messages on party $c2$ are interpreted as positive on party $c1$, and vice-versa.

\section{Methodology}\label{sec:methodology}

\begin{figure}[t]
\centering
    \includegraphics[width=0.8\textwidth]{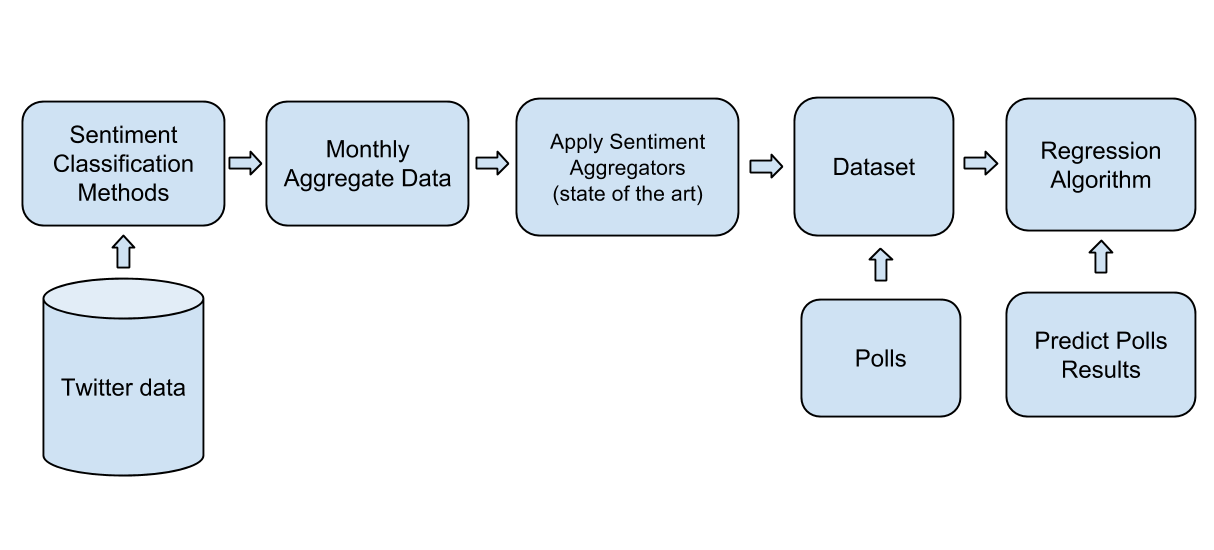}
\caption{Data mining process pipeline.}
\label{fig:architecture}
\end{figure}

Figure \ref{fig:architecture} depicts an overview of the data mining process pipeline applied in this work. To collect and process raw Twitter data, we use an online reputation monitoring platform \cite{saleiro2015popmine} which can be used by researchers interested in tracking political opinion on the web. It collects tweets from a pre-defined sample of users, applies named entity disambiguation \cite{saleiro2013popstar} and generates indicators of both frequency of mention and polarity (positivity/negativity) of mentions of entities across time. In our case, tweets are collected from the stream of 100 thousand different users, representing a sample of the Portuguese community on Twitter. 

The platform automatically classifies each tweet according to its sentiment polarity. If a message expresses a positive, negative or neutral opinion regarding an entity (e.g. politicians), it is classified as positive, negative or neutral mention, respectively. The sentiment classifier uses a corpus of 1500 annotated tweets as training set and it is reported an accuracy over 80\% using 10-fold cross validation. These 1500 tweets were manually annotated by 3 political science students. 

Mentions of entities and respective polarity are aggregated by counting positive, negative, neutral and total mentions for each entity in a given period. Sentiment aggregate functions use these cumulative numbers as input to generate a new value for each specific time period. Since we want to use sentiment aggregate functions as features of a regression model to produce an estimate of the political opinion, we decided to use traditional poll results as gold standard.

\subsection{Sentiment Aggregate Functions}

The following list presents the sentiment aggregate functions applied to the aggregated data between polls:

\begin{itemize}
\item[-]$entity\_buzz$: the monthly sum of the number of mentions (buzz) of a given entity (political party leader) between consecutive polls.

\item[-]$entity\_positives$: the monthly sum of the positively classified mentions of a given entity (political party leader) between consecutive polls.

\item[-]$entity\_neutrals$: the monthly sum of the neutral classified mentions of a given entity (political party leader) between consecutive polls.

\item[-]$entity\_negatives$: the monthly sum of the negatively classified mentions of a given entity (political party leader) between consecutive polls.

\item[-]$bermingham$ \cite{Bermingham}: $\log_{10}{\frac{entity\_posistives +1}{entitty\_negatives +1}}$

\item[-]$berminghamsovn$ \cite{Bermingham}: $\frac{entity\_negatives}{total\_negatives} $, $total\_negatives$ \\ \\ corresponds to the sum of the negative mentions of all entities between polls.

\item[-]$berminghamsovp$ \cite{Bermingham}: $\frac{entity\_positives}{total\_positives}$,  $total\_positives$ corresponds to the sum of the positives mentions of all entities between polls.

\item[-]$connor$ \cite{Connor}:  $\frac{entity\_positives}{entity\_negatives}$

\item[-]$gayo$ \cite{Gayo-Avello}: $\frac{entity\_positives + others\_negatives}{total\_positives + total\_negatives}$ 

\item[-]$polarity$:  $entity\_positives - entity\_negatives$

\item[-]$polarityONeutral$: $\frac{entity\_positives - entity\_negatives}{entity\_neutrals}$

\item[-]$polarityOTotal$: $\frac{entity\_positives - entity\_negatives}{entity\_buzz}$ 

\item[-]$subjOTotal$: $\frac{entity\_positives + entity\_negatives}{entity}$

\item[-]$subjNeuv$: $\frac{entity\_positives + entity\_negatives}{entity\_neutrals}$

\item[-]$subjSoV$: $\frac{entity\_positives + entity\_negatives}{total\_positives + total\_negatives }$

\item[-]$subjVol$: $entity\_positives + entity\_negatives$

\item[-]$share$ \cite{Bermingham}: $\frac{entity\_buzz}{total\-buzz}$

\item[-]$shareOfNegDistribution$: $\frac{ \frac{entity\_negatives}{entity\_buzz}}{\sum_{i=0}^n \frac{entity\_negatives_i}{entity\_buzz_i}}$, where $n$ is the number of political entities in the poll

\item[-]$normalized\_positive$: $\frac{entity\_positivesi}{entity\_buzz}$

\item[-]$normalized\_negative$: $\frac{entity\_positivesi}{entity\_buzz}$

\item[-]$normalized\_neutral$: $\frac{entity\_positivesi}{entity\_buzz}$

\item[-]$normalized\_bermingham$: $\log_{10}{\frac{normalized\_positives +1}{normalized\_negatives +1}} $

\item[-]$normalized\_connor$: $\frac{normalized\_positives}{normalized\_negatives}$

\item[-]$normalized\_gayo$: \\ \\
$\frac{normalized\_positives + normalized\_others\_negatives}{normalized\_total\_positives + normalized\_total\_negatives}$

\item[-]$normalized\_polarity$: \\ \\ $normalized\_positives - normalized\_negatives$

\end{itemize}

The sentiment aggregate functions are used as features in the regression models.

\section{Data}\label{sec:data}

The data used in this work consists of tweets mentioning Portuguese political party leaders and polls from August 2011 to December 2013. This period corresponds to the Portuguese bailout when several austerity measures were adopted by the incumbent right wing governmental coalition of PSD and CDS. 

\subsection{Twitter}
\begin{table}[!t]
\centering
\caption{Distribution of positive, negative and neutral mentions per political party}\label{tab:twitter_data_dist}
    \begin{tabular}{ l | r | r | r | r |}
    \cline{2-5}
     &Negative & Positive & Neutral & Total Mentions \\ \cline{2-5} \hline
     \multicolumn{1}{ |c| }{PSD} & 69 723 & 121 & 37 133 &106 977\\ \hline
     \multicolumn{1}{ |c| }{PS} & 28 660 & 225 & 15 326 & 44 211 \\ \hline
     \multicolumn{1}{ |c| }{CDS} & 41 935 & 51 & 17 554 & 59 540 \\ \hline
     \multicolumn{1}{ |c| }{CDU} & 2 445 & 79 & 5 604 & 8 128 \\ \hline
     \multicolumn{1}{ |c| }{BE} & 9 603 & 306 & 4 214 & 14 123 \\ \hline
    \end{tabular}
\end{table} 

The Twitter data set contains 232 979 classified messages, collected from a network of 100 thousand different users classified as Portuguese. Table \ref{tab:twitter_data_dist} presents the distribution of positive, negative, and neutral mentions of the political leaders of the 5 most voted political parties in Portugal (PSD, PS, CDS, PCP and BE).
The negative mentions represent the majority of the total mentions, except for CDU where the number of negative mentions is smaller than the neutral ones. The positive mentions represent less than 1\% of the total mentions of each party, except for BE where they represent 2\% of the total mentions. The most mentioned parties are PS, PSD and CDS. The total mentions of these three parties represent 90\% of the data sample total mentions.  

\begin{figure}[!t]
\centering
    \includegraphics[width=0.9\textwidth]{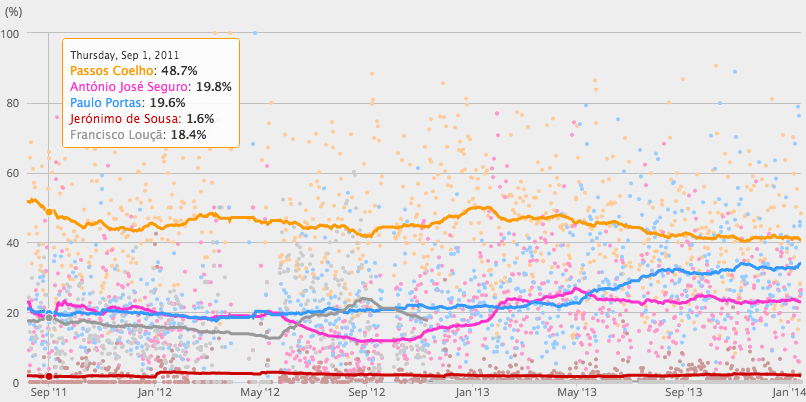}
\caption{Negatives share ($berminghamsovn$) of political leaders in Twitter.}
\label{fig:negshare}
\end{figure}

Figure \ref{fig:negshare} depicts the time series of the $berminghamsovn$ (negatives share) sentiment aggregate function. The higher the value of the function the higher is the percentage of negative tweets mention a given political entity in comparison with the other entities. As expected, Pedro Passos Coelho (PSD) as prime-minister is the leader with the higher score through the all time period under study. Paulo Portas (CDS) leader of the other party of the coalition, and also member of the government is the second most negatively mentioned in the period, while António José Seguro (PS) is in some periods the second higher.
PSD and CDS are the incumbent parties while PS is the main opposition party in the time frame under study.  PSD and CDS as government parties were raising taxes and cutting salaries. PS was the incumbent government during the years that led to the bailout and a fraction of the population considered responsible for the financial crisis. The bailout and the consequent austerity measures could explain the overwhelming percentage of negative mentions although we verified that in other time periods the high percentage of negatives mentions remains. We can say that Twitter users of this sample when mentioning political leaders on their tweets tend to criticize them.

\subsection{Political Opinion Polls}

\begin{figure}[!t]
\centering
    \includegraphics[width=0.7\textwidth]{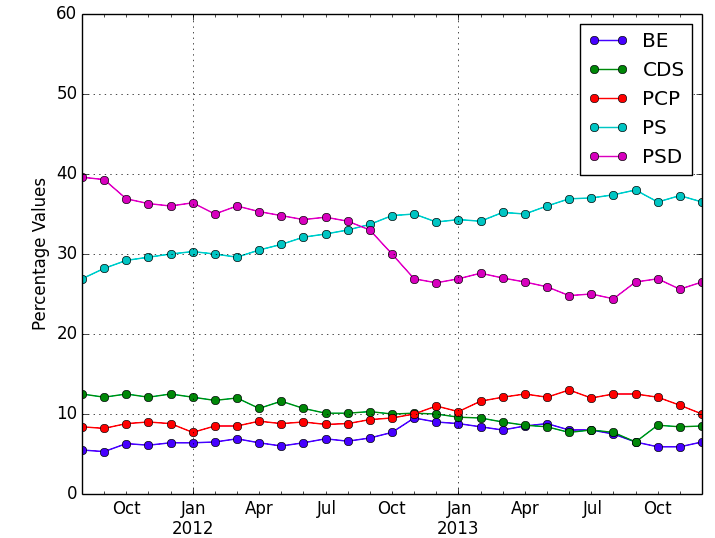}
\caption{Representation of the monthly poll results of each political candidate}
\label{fig:pollEuro}
\end{figure}

The polling was performed by Eurosondagem, a Portuguese private company which collects public opinion. This data set contains the monthly polls results of the five main Portuguese parties, from June 2011 to December 2013. Figure \ref{fig:pollEuro} represents the evolution of Portuguese polls results. We can see two main party groups: The first group, where both PSD and PS are included, has a higher value of vote intention (above 23\%). PSD despite starting as the preferred party in vote intention, has a downtrend along the time, losing the leadership for PS in September 2012. On the other hand, PS has in general an uptrend. The second group, composed by CDS, PCP and BE, has a vote intention range from 5\% to 15\%. While CDS has a downtrend in public opinion, PCP has an ascendent one. Although the constant tendencies (up- and downtrends), we noticed that the maximum variation observed between two consecutive months is 3\%. 
In June 2013 there was political crises in the government when CDS threaten to leave the government coalition due to the austerity measures being implemented and corresponds to the moment when PS takes the lead in the polls.

\section{Experimental Setup}\label{sec:experimental}
{\renewcommand{\arraystretch}{2}


We defined the period of 2011 to December 2012 as training set and the all year of 2013 as test set. We applied a sliding window setting in which we start to predict the poll results of January 2013 using the previous 16 months as training set. The second poll to estimate is February 2013 and we train a new model using the previous 16 months to the target month under prediction.

\begin{itemize}
\item Training set – containing the monthly values of the aggregators (both sentiment and buzz aggregator) for 16 months prior the month intended to be predicted. 
\item Testing set - containing the values of the aggregators (both sentiment and buzz aggregator) of the month intended to be predicted.
\end{itemize}

\begin{enumerate}
\item We select the values of the aggregators of the 16 months prior January 2013 (September 2011 to December 2012). 
\item We use that data to train our regression model.
\item Then we input the aggregators' values of January 2013 - the first record of the testing set - in the the trained model, to obtain the poll results prediction. 
\item We select the next month of the testing set and repeat the process until all months are predicted.
\end{enumerate}

The models are created using two regression algorithms: a linear regression algorithm (Ordinary Least Squares - OLS) and a non-linear regression algorithm (Random Forests - RF). 
We also run an experiment using the derivative of the polls time series as gold standard, i.e., poll results variations from poll to poll. Thus, we also calculate the variations of the aggregate functions from month to month as features. 
Furthermore, we repeat each experiment including and excluding the lagged self of the polls, i.e., the last result of the poll for a given candidate ($y_{t-1}$)  or the last polls result variation ($\Delta y_{t-1}$) when predicting polls variations.
We use Mean Absolute Error (MAE) \cite{Han} as evaluation measure, to determine the absolute error of each prediction. Then, we calculate the average of the twelve MAE's so we could know the global prediction error of our model. 

\begin{equation}
MAE = \frac{\sum_{i=1}^{n}|f_{i} - y_{i}|}{n}
\label{eq:mae}
\end{equation}
$n$ is the number of forecasts, $f_{i}$ is the model's forecast and $y_{i}$ the real outcome.

\section{Results and Discussion}\label{sec:results}
In this Section we explain in detail the experiments and its results. We perform two different experiments: (1) using absolute values and (2) using monthly variations.

\subsection{Predicting Polls Results}
In this experiment, the sentiment aggregators take absolute values in order to predict the absolute values of polls results. Mathematically speaking, this experiment can be seen as:
$y \leftarrow$ \{$y_{t-1}$, $buzzAggregators$, $sentimentAggregators$\}. In figure \ref{fig:expabs_mae} we see the global errors we obtained.

\begin{figure}[!t]
\centering
    \includegraphics[scale = 0.40]{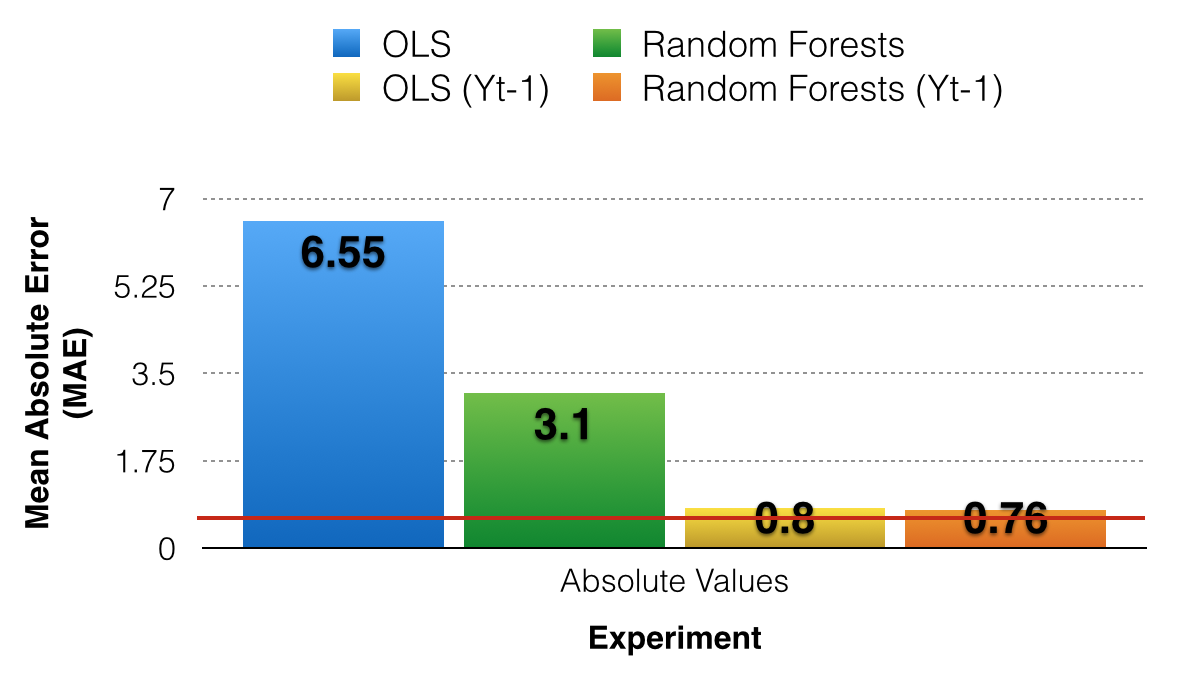}
\caption{Error predictions for polls results.}
\label{fig:expabs_mae}
\end{figure}

The results shows that we obtain an Mean Absolute Error for the 5 parties poll results over 12 months of 6.55 \% using Ordinary Least Squares and 3.1 \% using Random Forests. The lagged self of the polls, i.e., assuming the last known poll result as prediction results in a MAE of 0.61 \% which was expectable since the polls exhibit slight changes from month to month. 
This experiment shows that the inclusion of the lagged self ($y_{t-1}$) produces average errors similar to the lagged self.

\subsection{Predicting Polls Results Variation}
According to our exploratory data analysis, the polls results have a small variation between two consecutive months. Thus, instead of predicting the absolute value of poll results, we tried to predict the variation, $\Delta y \leftarrow $ \{$\Delta (y_{t-1})$, $\Delta buzzAggregators$, $\Delta sentimentAggregators$\}

\begin{figure}[!t]
\centering
    \includegraphics[scale = 0.30]{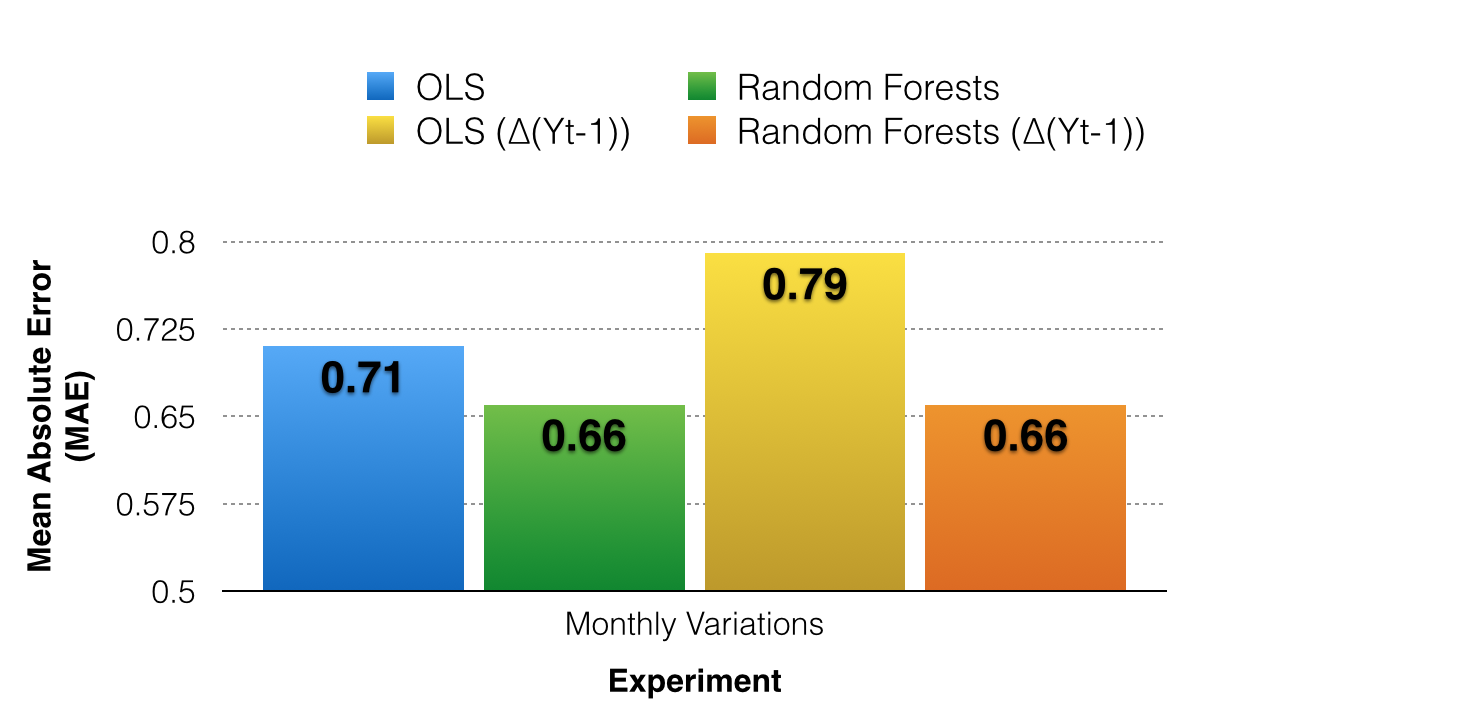}
\caption{Error predictions for polls results variation.}
\label{fig:exp_var_mae}
\end{figure}

In this particular experiment, the inclusion of the $\Delta y_{t-1}$ as feature in the regression model has not a determinant role (figure \ref{fig:exp_var_mae}). Including that feature we could not obtain lower MAE than excluding it. It means that the real monthly poll variation is not constant over the year. In general, using a non-linear regression algorithm we obtain lower MAE. The results show that when leading with polls results with slight changes from poll to poll it makes sense to transform the dataset by derivation.

\subsubsection{Buzz and Sentiment}
Several studies state that the buzz has predictive power and reflects correctly the public opinion on social media. Following that premise, we trained our models with buzz and sentiment aggregators separately to predict polls variations:
\begin{itemize}
\item $\Delta y \leftarrow$ \{$\Delta (y_{t-1})$, $\Delta buzzAggregators$\}
\item $\Delta y \leftarrow$ \{$\Delta (y_{t-1})$, $\Delta sentimentAggregators$\}
\end{itemize}
This experiment allowed us to compare the behavior of buzz and sentiment aggregators.

\begin{figure}[Ht]
\centering
    \includegraphics[scale = 0.30]{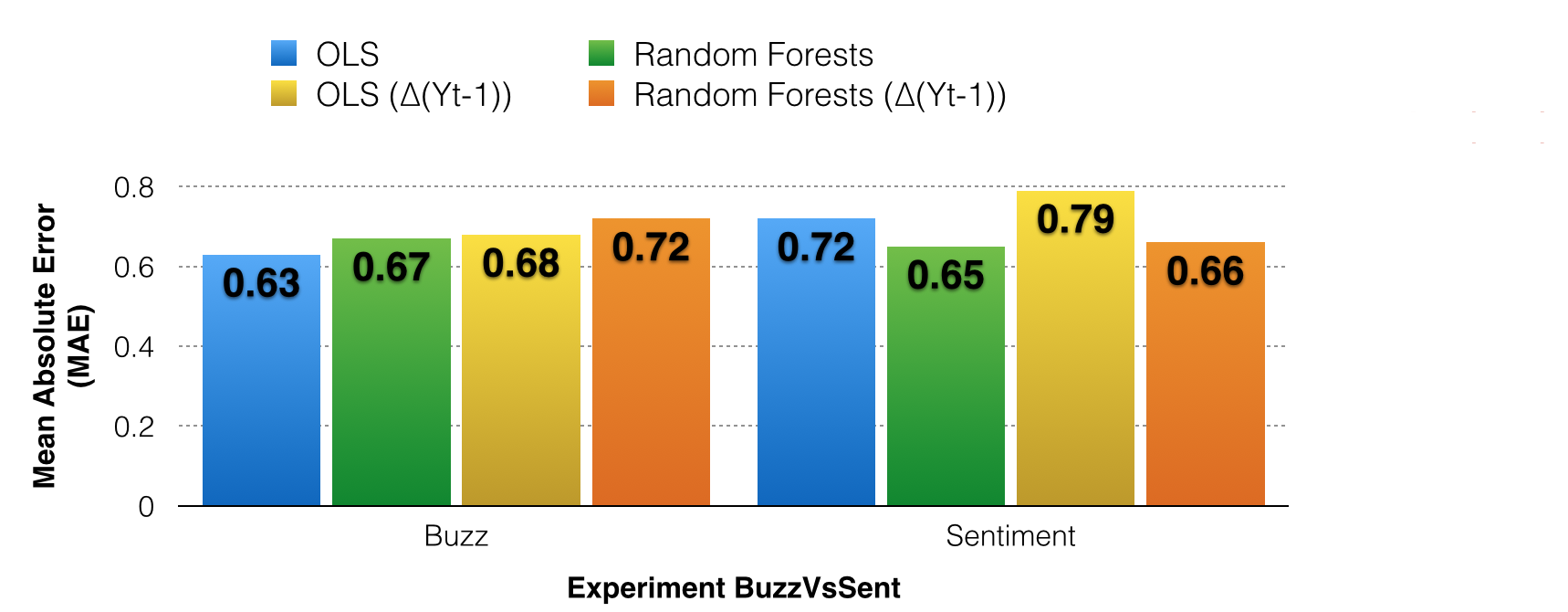}
\caption{Mean absolute error buzz vs sentiment.}
\label{fig:buzzvssent}
\end{figure}

According to figure \ref{fig:buzzvssent}, buzz and sentiment aggregators have similar results. Although the OLS algorithm combined only with buzz aggregators has a slightly lower error than the other models, it is not a significant improvement. These results also show that Random Forests algorithm performs the best when combined only with sentiment aggregators.

\subsection{Feature Selection}
One of the main goals of our work is to understand which aggregator (or group of aggregators) better suits our case study. According to the previous experiments, we can achieve lower prediction errors when training our model with buzz and sentiment aggregators separately. However, when training our model with these two kinds of aggregators separately, we are implicitly performing feature selection. We only have two buzz features ($share$ and $total\_mentions$). Due to that small amount of features, it was not necessary to perform any feature selection technique within buzz features. Thus, we decided to apply a feature selection technique to the sentiment aggregators, in order to select the most informative ones to predict the monthly polls results variation. We use univariate feature selection, selecting 10\% of the sentiment features (total of 3 features). Using this technique, the Random Forests' global error raise from 0.65 to 0.73. However, OLS presents an MAE drop from 0.72 to 0.67. Another important fact to notice is that if we perform univariate feature selection to all aggregators (buzz and sentiment), we will achieve the same MAE value that when applied only to sentiment aggregators. 

We try a different approach and perform a recursive feature elimination technique. In this technique, features are eliminated recursively according to a initial score given by the external estimator. This method allow us to determine the number of features to select. Thus, also selecting 3 features, the OLS' MAE drop to 0.63. Once again, none of the buzz features were selected. 
Furthermore, both feature selection techniques select different features for each monthly prediction.

\begin{figure*}[t]
\centering
    \includegraphics[width=0.9\textwidth]{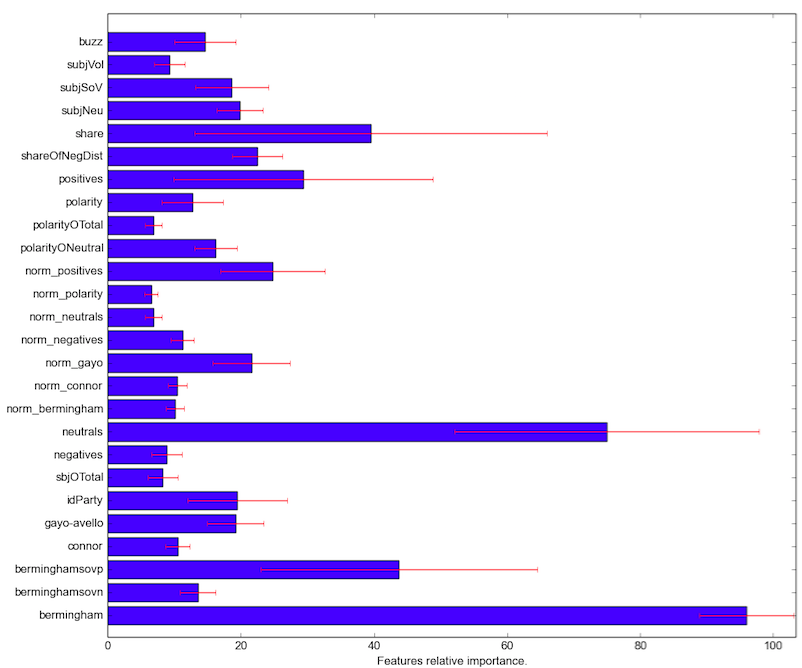}
\caption{Aggregate functions importance in the Random Forests models.}
\label{fig:importances}
\end{figure*}
\subsection{Feature Importance}

We select the Random Forest model of monthly variations to study the features importance as depicted in figure \ref{fig:importances}. 
The higher, the more important the feature. The importance of a feature is computed as the (normalized) total reduction of the criterion brought by that feature. It is also known as the Gini importance. Values correspond to the average of the Gini importance over the different models trained in the experiments. The single most important feature is the $bermingham$ aggregate function, followed by $neutrals$. It is important to notice that when combining all the aggregate functions as features in a single regression model, the $buzz$ does not comprises a high Gini importance, even if when used as a single feature produces similar results with the sentiment aggregate functions. In general, the standard deviation of the GIni importance is relatively high. This has to due with our experimental setup, as the values depicted in the bar chart correspond to the average of the Gini importance over 12 different models (12 months of testing set). Therefore, feature importances vary over time while the MAE tends to remain unchanged. We can say that different features have different informative value over time and consequently it is useful to combine all the sentiment aggregation functions as features of the regression models over time.

\section{Conclusions}\label{sec:conclusions}

We studied a large set of sentiment aggregate functions to use as features in a regression model to predict political opinion poll results. The results show that we can estimate the polls results with low prediction error, using sentiment and buzz aggregators based on the opinions expressed on social media. We introduced a strong baseline for comparison, the lagged self of the polls.
In our study, we built a model where we achieve the lowest MAE using the linear algorithm (OLS), combined only with buzz aggregators, using monthly variations. The model has an MAE of 0.63\%.
We performed two feature selection techniques: (1) Univariate feature selection and (2) recursive feature elimination. Applying the recursive technique to the sentiment features, we can achieve an MAE of 0.63, equating our best model. Furthermore, the chosen features are not the same in every prediction.
Regarding feature importance analysis our experiments showed that $bermingham$ aggregate function represents the higher Gini importance in the Random Forests model.
The next immediate step is to implement a methodology using time series analysis. Furthermore, it is desirable to test this methodology with difference data sources, such as Facebook messages, blogs or news.
Other alternative approach we intend to implement and evaluate is to interpret this problem as a classification problem - predict only the changing direction of opinion poll result (i.e., up or down).

%
\bibliographystyle{unsrt}
\bibliography{refs}  
%
%
\end{document}